\documentclass[preprint,english,aps,prl,superscriptaddress,showpacs]{revtex4}
\usepackage{amsmath,amssymb}
\usepackage{graphicx}
\usepackage{babel}

\begin{document}

\title{Peak-effect and surface crystal-glass transition for surface-pinned
vortex array}

\author{B. Pla\c{c}ais}\email{placais@lpa.ens.fr}
\affiliation{Laboratoire Pierre Aigrain, D{\'e}partement de
Physique de l'Ecole Normale Sup\'erieure, 24 rue Lhomond, 75231
Paris Cedex 05, France}
\author{N. L\"utke-Entrup}
\affiliation{Laboratoire Pierre Aigrain, D{\'e}partement de
Physique de l'Ecole Normale Sup\'erieure, 24 rue Lhomond, 75231
Paris Cedex 05, France}
\author{J. Bellessa}
\affiliation{Laboratoire Pierre Aigrain, D{\'e}partement de
Physique de l'Ecole Normale Sup\'erieure, 24 rue Lhomond, 75231
Paris Cedex 05, France}
\author{P. Mathieu}
\affiliation{Laboratoire Pierre Aigrain, D{\'e}partement de
Physique de l'Ecole Normale Sup\'erieure, 24 rue Lhomond, 75231
Paris Cedex 05, France}
\author{Y. Simon}
\affiliation{Laboratoire Pierre Aigrain, D{\'e}partement de
Physique de l'Ecole Normale Sup\'erieure, 24 rue Lhomond, 75231
Paris Cedex 05, France}
\author{E.B. Sonin}
\affiliation{Racah Institute of Physics,  Hebrew University of
Jerusalem - Jerusalem 91904, Israel}

\begin{abstract}
We present a theoretical and experimental study of the peak effect
in the surface pinning of vortices. It is associated with a sharp
transition in the vortex slippage length which we relate to a
crossover from a weakly disordered crystal to a surface glass
state. Experiments are performed on ion-beam etched Nb crystals.
The slippage length is deduced from 1kHz-1MHz linear AC
penetration depth measurements.
\end{abstract}


\maketitle

A peak in the {\em critical current vs magnetic field} plot, the
peak-effect (PE),  is observed in superconductors close to the
transition line where the critical current vanishes. From the very
first studies \cite{Autler62,Kes81,Pippard69,Larkin} it was
supposed that PE originates from softening of the vortex lattice
(VL) by disorder near the transition. This results in a more
effective vortex pinning, which corresponds to a higher critical
current. The phenomenon is directly connected with a fundamental
problem of the condensed matter physics: the competition between
elasticity and disorder. Numerous scenarios of PE have been
discussed, but all of them dealt with the competition between
vortex elasticity and {\em bulk} pinning. Here we present an
essentially different scenario of PE: it is {\em surface} pinning
of vortices which interplays with {\em bulk} vortex elasticity.

\begin{figure}[!htbp]
        \centerline{\includegraphics[width=5in,
  keepaspectratio]{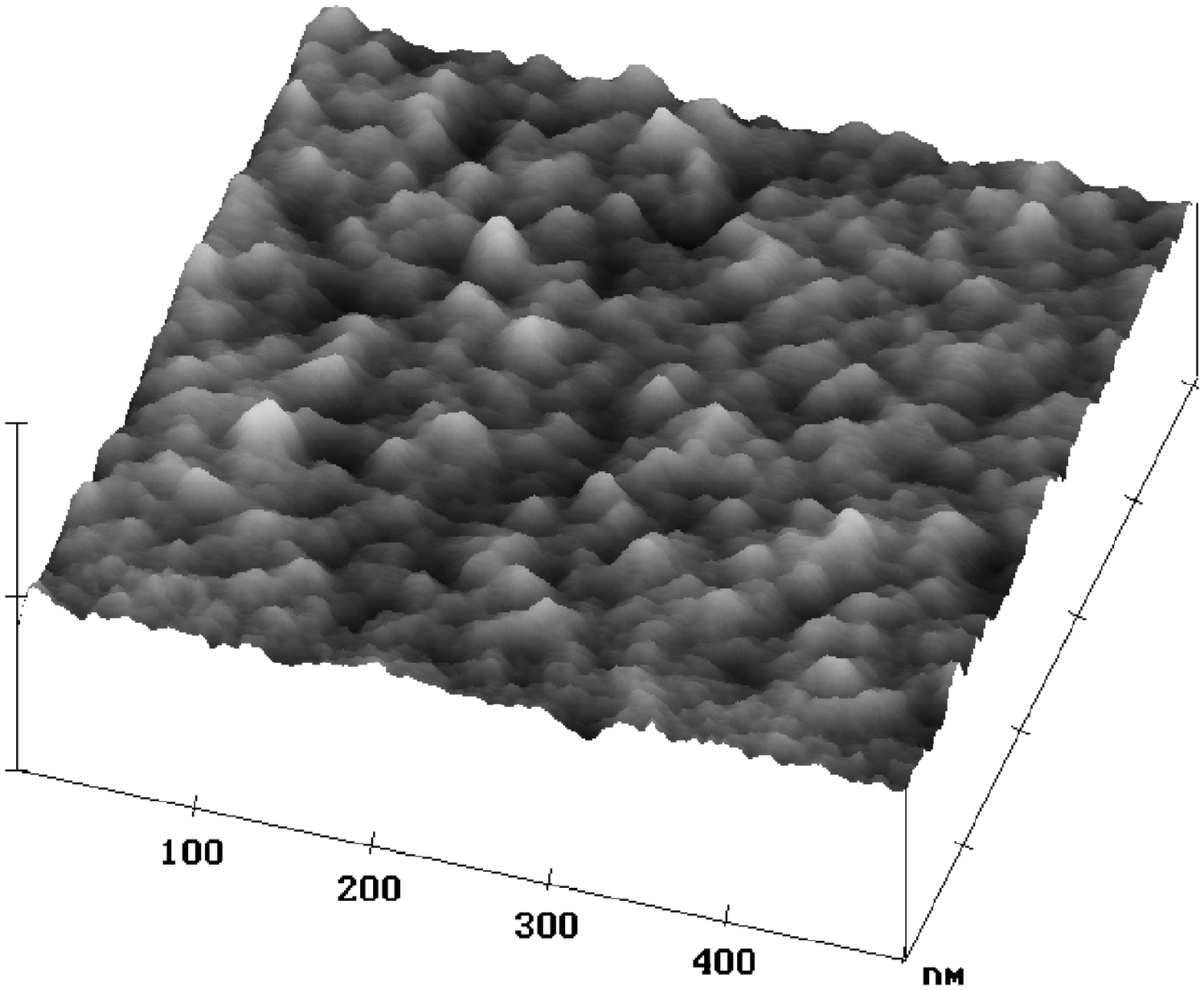}}
        \caption{Atomic-force
microscopy (AFM) picture of the etched Nb surface.}\label{AFM1}
        \end{figure}

\begin{figure}[!htbp]
        \centerline{\includegraphics[width=5in,
  keepaspectratio]{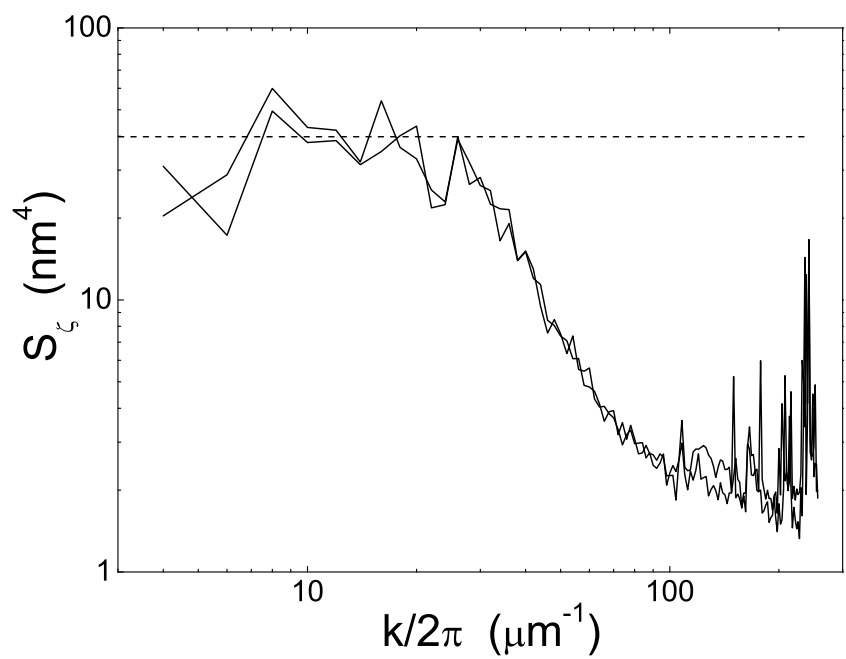}}
        \caption{Spectrum of the surface corrugation. The dashed line is a
guide.  The cut-off at $\sim 40\mu$m$^{-1}$ is due to AFM tip
diameter.} \label{AFM2}
        \end{figure}

A controlled surface roughness $\zeta({\mathbf r})$ is obtained
 by etching the Nb sample surfaces with 500eV-Ar$^+$ ions (Fig.\ref{AFM1}).
The sputtering of Nb atoms by low energy ions is a stochastic
process. It gives rise to a white corrugation spectrum
$S_{\zeta}({\mathbf k})=\int {\mathbf dr}\,e^{-i\mathbf k \mathbf
r}\langle \zeta({\mathbf r}+{\mathbf R})\zeta({\mathbf
R})\rangle_{\mathbf R} \simeq a^3\overline{\Delta z}/\pi$ for
$|{\mathbf k}|< 1/a$, where $a=0.26$ nm is the  Nb lattice
parameter and $\overline{\Delta z}$ the average etching depth. In
our experiment $\overline{\Delta z}\sim10\mu$m for a 90 min
exposure to a 1.5 mA/cm$^2$ Ar$^+$ flux  so that $S_{\zeta}\sim 50
$ nm$^4$ and the  total roughness $\zeta^*=\sqrt{\langle
\zeta({\mathbf r})^2\rangle}<(a\overline{\Delta z})^{1/2}= 50$ nm.
Atomic force microscopy (AFM) in Fig.\ref{AFM2} confirms the above
estimates with $S_{\zeta}( k)\simeq 40$ nm$^{4 }$ for $k\lesssim40
\mu$m$^{-1}$.  Unfortunately, the finite AFM-tip radius masks the
large $k$ spectrum, so that we can only bracket the upper cut-off
$k_c$ of $S_{\zeta}(k)$ in the range $10^{-2}<k_ca<1$. This
entails a large uncertainty in $\zeta^*=0.5$--$50$ nm.
Importantly, AFM data indicate the presence of roughness at small
scale, with wave numbers $k\sim a_0^{-1}\simeq\! (50$nm)$^{-1}$
where $a_0=\sqrt{\pi B/\varphi_0}$ is the VL reciprocal unit and
$\varphi_0=h/2e$ is the flux quantum.

The peak effect is generally observed in the critical current data
$I_c(B)$ or $I_c(T)$. Since $I_c\gtrsim 20$ A are quite large in
our Nb samples, we prefer to rely on the AC linear surface
impedance $Z(\omega)=-i\omega \mu_0 \lambda_{AC}$ which is a more
accurate probe of the vortex state, especially in the vicinity of
a transition. According to \cite{Entrup1,Entrup2}, the AC
penetration depth $\lambda_{AC}$ in thick samples is given by
\begin{equation}
\frac{1}{\lambda_{AC}}=\frac{1}{L_S}+\left(\frac{1}{\lambda_{C}^2}+i\omega
\mu_0\sigma_f\right)^{1/2};\quad
L_S=\frac{l_SB}{\mu_0\varepsilon}\quad .
 \label{two_mode}\end{equation}
Here $\sigma_f$ is the flux-flow resistivity,
$\varepsilon\varphi_0$ is the vortex-line tension, and $\lambda_C$
the Campbell  depth for bulk pinning \cite{Campbell}. Expression
(\ref{two_mode}) deviates from the Coffey-Clem theory \cite{CC} by
the addition of a surface pinning term $1/L_s$.  The surface
pinning length $L_S\!\sim\!0.1$--$100 ~\mu$m can indeed simulate a
Campbell length at low frequency but gives a very different
behaviour at finite frequency \cite{Entrup1,Entrup2}. The above
expression was derived within the frame of the {\em two-mode
electrodynamics} \cite{Entrup1,Sonin92}, which incorporates the
surface pinning by introducing a phenomenological boundary
condition,
\begin{equation}
\varepsilon\varphi_0\left({{\mathbf u} \over l_S}+  {\partial
{\mathbf u} \over \partial z}\right)=0~,
     \label{BC}\end{equation}
imposed on the VL displacement ${\mathbf u}(z)$  at the surface of
the sample, which occupies the semi-space $z<0$. Here $l_S$ is a
{\em slippage length} and the displacement ${\mathbf u}$ is
averaged over the position vectors ${\mathbf r}$ in the $xy$
plane. Equation (\ref{BC}) represents the balance between the
pinning force $-\varepsilon\varphi_0 {\mathbf u} / l_S$ and the
line tension force $\varepsilon\varphi_0 \partial {\mathbf u}
/\partial z$.

\begin{figure}[!htbp]
        \centerline{\includegraphics[width=5 in,
  keepaspectratio]{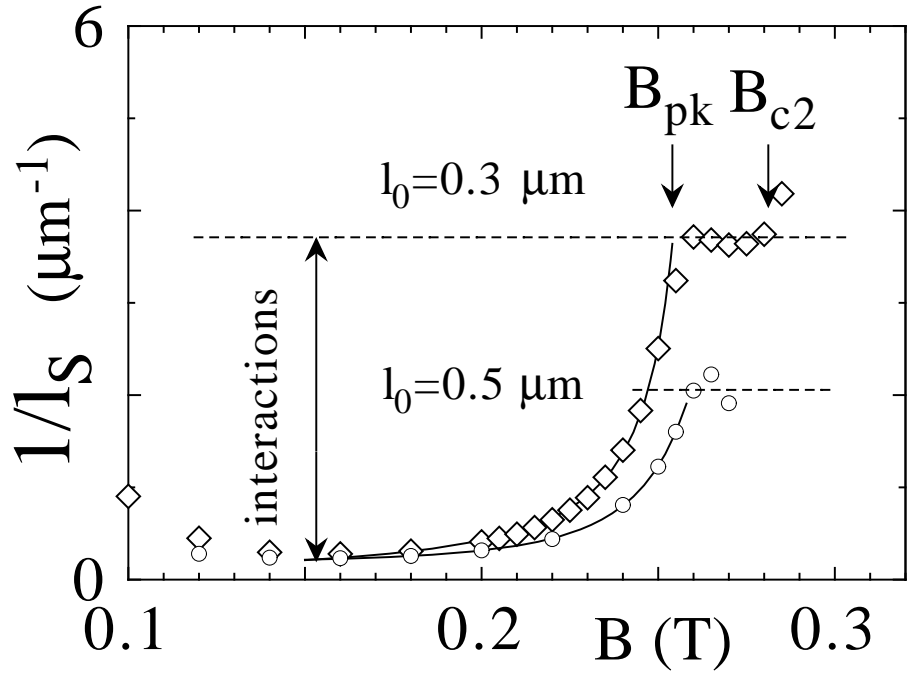}}
        \caption{The slippage length $l_S(B)$ for a vortex array at a
rough surface. It is deduced from the $L_S$ data  according to
$l_S=L_S\mu_0\varepsilon/B$ in Eq.(\ref{two_mode}). The effect of
vortex interactions in the collective-pinning regime below
$B_{pk}$ is visible as a suppression of $1/l_S(B)$. Solid lines
are power-law fits.} \label{slippage}
        \end{figure}

If vortices do not interact, the slippage length $l_S$ does not
depend on vortex density and is on the order of a curvature radius
of the surface profile (individual pinning). But in general $l_S$
may depend on vortex density, i.e. on magnetic field (see
Fig.\ref{slippage}). If vortices strongly interact the theory of
collective pinning \cite{Larkin} assumes that within the so-called
Larkin-Ovchinnikov domain of size $L_c$ the vortices move mostly
coherently without essential deformation of the vortex lattice.
But then because of the random directions of  pinning forces on
every vortex, the total force on vortices in the domain is
proportional  to $\sqrt{N_c}$ and not to $N_c=L_c^2/a_0^2$, the
number of vortices in the domain. Correspondingly the pinning
force per vortex must be smaller by the factor
$\sqrt{N_c}=L_c/a_0$, i.e. $1/l_S =1/l_0 \sqrt{N_c}=a_0/l_0 L_c$.
$L_c$ is  usually  derived  from the balance between the elastic
and pinning energy.  Pinning is collective as long as $N_c\gg1$.
The condition $N_c \sim 1$ (or $l_S \sim l_0$) determines the
crossover from the collective to the individual pinning. Later in
the paper we shall derive $l_S$ without these heuristic arguments.

\begin{figure}[!htbp]
        \centerline{\includegraphics[width=5in,
  keepaspectratio]{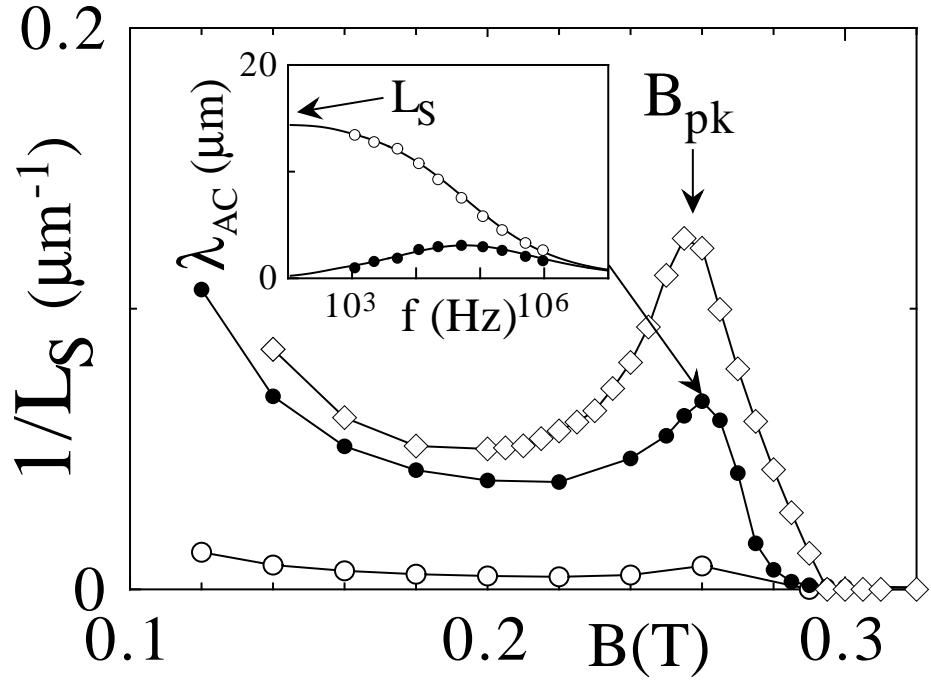}}
        \caption{Peak-effect in the elastic response $L_S^{-1}(B)$ of a
surface-pinned vortex array at T=4.2K. Diamonds and  full circles
correspond to the 45 degree and perpendicular field orientations.
Empty circles are the pristine sample measurement. \emph{Inset}:
the frequency dependence of the real and imaginary parts (open and
full circles respectively) of the penetration depth
$\lambda_{AC}(B,f)$ from which $L_S$ is deduced. Solid lines are
theoretical fit to Eq.(\ref{two_mode}) with $\lambda_C=\infty$,
$\sigma_f^{-1}=10$nOhm.cm and $L_S=14.9\mu$m.}\label{pic_effect}
        \end{figure}

Figure \ref{pic_effect} shows the PE in the inverse
surface-pinning length. The sample, with dimensions
$25\!\times\!10.1\!\times\!0.87$ mm$^3$, was annealed in
ultra-high vacuum which gives a low residual resistivity
$\rho_{\rm n}\!=\!11$ n$\Omega$cm (resistivity ratio $\sim1300$)
and an upper critical field $B_{\rm c2}\!=\!0.29$ T at $4.2 K$
\cite{Williamson70}. Data points are obtained by fitting the
penetration-depth spectra (inset of the figure) with
Eq.(\ref{two_mode}). Metastability in the vortex density ($\pm
0.005$ T) and/or arrangement is removed by feeding a large
transient current $I>>I_c$ in the sample prior to measurement. The
abrupt onset of the AC-flux penetration along the samples edges
which are parallel to the field precludes quantitative analysis
for $B\gtrsim 0.95 B_{c2}$; this  difficulty is overcome by
working in oblique field (diamonds in Fig.\ref{pic_effect}).
Already present in the pristine sample, the PE is strongly
enhanced in ion-etched samples (circles in Fig.\ref{pic_effect}).
By contrast chemically-etched samples show little PE but a large
increase  of pinning at lower fields. We think that this
difference is due to the lack of small-scale corrugation in the
wet-etching techniques.

We quantitatively separate the bulk and surface pinning
contributions, $\lambda_C$ and $L_S$, by fitting the full
1kHz--1MHz spectrum $\lambda_{AC}(f)$ with Eq.(\ref{two_mode}).
Remarkably we always find that $\lambda_C$ is much larger than the
sample thickness ($\sim 1$ mm), whereas $L_S\sim 5$--$100\mu$m,
meaning that bulk pinning is negligible. This observation, which
is true for $\lambda_{AC}(f)$ spectra taken on both sides of the
peak, confirms that surface pinning is most relevant in our
experiment. The oblique-field (45 degree) data are larger by a
factor $\sim\!2$; this is due to surface-reinforcement of
superconductivity in tilted fields. Otherwise, data are similar at
lower temperatures with however larger $B_{pk}\!=\!0.95B_{c2}$
(1.8K) and a less pronounced peak resembling sometimes to a
shoulder.

Using the Abrikosov expression \cite{Abrikosov},
$\mu_0\varepsilon\!\simeq\!(B_{c2}\!-\!B)/2.32\kappa^2$ with
$\kappa\!=\!\lambda/\xi\!=\!1.3$ ($\xi$ is the coherence length
and $\lambda$ is the London penetration depth), we deduce from
Eq.(\ref{two_mode}) the $l_S(B)$-data  in Fig.\ref{slippage}. The
high-field plateaus $l_S(B)$ above $B_{pk}$ are suggestive of
individual pinning, when $l_S \sim l_0$ does not depend on $B$.
Note that value of the contact angle for VL at the surface,
$a_0/l_0\!\simeq\!0.1$ estimated from $l_0\!\simeq\!0.5~\mu$m
(normal field) and $a_0=50$ nm, fits in the window
$0.01<\zeta^*/a_0<1$ prescribed by corrugation geometry. By
contrast, the strong suppression of $1/l_S$ below $B_{pk}$ (factor
$\sim\!10$ in oblique field) reflects the {\em collective} regime
of surface pinning, which was known earlier in rotating $^3$He
\cite{KKKS}. The transition is sharp unlike the continuous ones
reported in Refs.\cite{Paltiel00,Ling01}. Thus the experiment
provides an evidence that PE is accompanied by the crossover from
collective to individual surface pinning. In the following, we
give the theory for the slippage length  $l_S(B)$ in the
collective regime and explain its vanishing at the PE transition.

The first step of our analysis addresses the response of the
semi-infinite VL to a Fourier component ${\mathbf f}({\mathbf r})
= {\mathbf f}({\mathbf  k})e^{i {\mathbf k  r}}$ of the surface
force on vortices. The force produces vortex displacements in the
sample bulk ($z<0$) in the form ${\mathbf u}({\mathbf r},z) =\sum
\limits_{k_z}{\mathbf U}({\mathbf k}, k_z) e^{i{\mathbf k}
{\mathbf r}+i k_z z}$. We look for the elastic constant
$C({\mathbf  k})= f({\mathbf k})/u({\mathbf  k})$, connecting the
Fourier components of the surface force ${\mathbf f}({\mathbf
 k})$ to the surface displacement ${\mathbf u}({\mathbf
k})=\sum \limits_{k_z}{\mathbf U}({\mathbf k}, k_z)$. The force is
assumed to be transverse [${\mathbf f}({\mathbf  k})$, ${\mathbf
U}({\mathbf k},k_z) \perp \mathbf k$], since VL compressibility is
quite low and the response to the longitudinal force is weak. The
possible values of out-of-plane wave-vector component $k_z$ must
be found from the equation of the elasticity theory:
\begin{equation}
\left[C_{66} k^2 + C_{44} ( {\mathbf k},k_z) k_z^2\right]{\mathbf
U}({\mathbf k},k_z)=0~,
   \label{EE}\end{equation}
where $C_{66}$ is the shear modulus and
\begin{equation}
C_{44}({\mathbf k},k_z) = {B^2 \over \mu_0}\frac{1}{1 +
\lambda^2(k^2 +k_z^2)} +{\varepsilon B}    \end{equation} is the
tilt-modulus, which takes into account nonlocal effects due to
long-range vortex-vortex interaction. The general solution of
Eq.(\ref{EE}) is a superposition of {\em two evanescent modes} in
the bulk, ${\mathbf u}({\mathbf r},z) =  e^{i{\mathbf k} {\mathbf
r}}\left[{\mathbf U}({\mathbf k}, p_1) e^{p_1 z} +{\mathbf
U}({\mathbf k}, p_2) e^{p_2 z}\right]$ with $p_1 \approx
k\sqrt{C_{66}/\varepsilon B}\ll 1/\lambda$ and $p_2 \approx
1/\lambda\sqrt{B/\mu_0\varepsilon}\gg 1/\lambda$. In order to
determine the two amplitudes ${\mathbf U}({\mathbf k}, p_1) $ and
${\mathbf U}({\mathbf k}, p_2) $, we need a second boundary
conditions.  It is imposed on the magnetic field, which is
determined from the London equation, ${\mathbf h}({\mathbf k},
k_z)=i k_z B {\mathbf U}({\mathbf k},k_z)\left[1+\lambda^2( k^2 +
k_z^2)\right]^{-1}$, and should vanish at the sample border
(transverse waves). Eventually this yields for $C_{66} \ll
\varepsilon B \ll B^2/\mu_0$:
\begin{equation}
C(k) \approx k\varphi_0\sqrt{{C_{66}\over \mu_0} {(1+ \lambda
^2k^2 \mu_0\varepsilon/B)\over 1+\lambda ^2k^2}}~.
 \label{CdeK}  \end{equation}
At large $\lambda k$ Eq.(\ref{CdeK}) gives
$C(k)=k\varphi_0\sqrt{\varepsilon C_{66}/ B}$.

The second step consists in calculating the deformations produced
by surface pinning from the corrugation profile $\zeta({\mathbf
r})$. The random force on the vortices is $ {\mathbf f}({\mathbf
r}_i)=-\varepsilon\varphi_0 {\mathbf\nabla }\zeta({\mathbf r}_i)$,
where ${\mathbf r}_i$ is the 2D position vector of the $i$th
vortex. The Fourier component of the force is ${\mathbf
f}({\mathbf  k})\!=\!-\!\varepsilon B \sum\limits _{\mathbf  Q}i
 \left[ {\mathbf  Q} - {\hat{\mathbf k}}{ (\hat{\mathbf  k}\cdot
{\mathbf Q}})\right] \int d {\mathbf r}\, e^{-i ({\mathbf k}
+{\mathbf  Q}){\mathbf r} }\zeta(\mathbf r )$, where the factor in
brackets separates the transverse component of the force and  the
summation over the reciprocal VL vector ${\mathbf Q}$ appears
because the force is applied in discrete sites of the VL.
Collecting contributions from all Fourier components ${\mathbf
u}({\mathbf  k})={\mathbf f}({\mathbf  k})/C({\mathbf  k})$ we
obtain the mean-square-root shear deformation at the surface
\begin{equation}
\left\langle{ ({\mathbf \nabla u})^2}\right\rangle =\left\langle
\left({\partial u_{x}\over \partial y} + {\partial u_{y}\over
\partial x}\right)^2\right\rangle  = {\varepsilon
^2\varphi_0^2\over 4\pi ^2 }\int  { k^2\,{\mathbf dk} \over
C(k)^2} \sum _{{\mathbf Q}}\left( Q^2 - { ({\mathbf k\cdot
Q})^2\over k^2}\right) S_\zeta({\mathbf k} +{\mathbf Q})~.
  \end{equation}
Here the integration over $\mathbf k$ is fulfilled over the VL
Brillouin zone. In the following we shall approximate the surface
corrugation spectrum by $S_\zeta({\mathbf k}) =2\pi \zeta^{*2}
r_d^{2}e^{-kr_d} $, where the corrugation correlation radius $r_d$
is determined by the spectrum cut-off $k_c$, if $k_c\xi <1$: $r_d
\sim k_c^{-1}$. But since the vortex cannot probe corrugation on
scales less than its ``size'' $\xi$, $r_d \sim \xi$ if $k_c\xi
>1$. Approximating the sum over ${\mathbf Q} $ by an integral, we
obtain for $k \sim 1/a_0 \ll Q\sim 1/r_d$:
\begin{equation}
\langle({\mathbf \nabla u})^2\rangle = {\varepsilon^2\varphi_0^2
a_0^2\over 8\pi}\int \limits_0^{2/a_0} { k^3\,d{ k} \over C(k)^2}
\int_0^\infty S_\zeta(Q) Q^3\,dQ \approx{\varepsilon B\over
C_{66}}{3r_d^2 \over l_0^2}. \label{def}\end{equation}
 Here we
used the expression $C(k)\approx k\varphi_0(C_{66}\varepsilon/
B)^{1/2}$ for large $k$, which is a good approximation when
$\lambda\gg a_0$. Note that since $C(k) \propto k$ at small $k$
the integral for the mean-square-root displacement $\langle
{\mathbf u}^2\rangle$ is divergent. This means that even a weak
disorder destroys the long range order near the surface as was
revealed in Ref. \cite{Feldman02}. However our analysis shows that
destruction of long-range order near the surface is not essential
for the peak effect, which is governed by the mean-square-root
deformation, but not by the mean-square-root displacement.

In the third step  we derive the boundary condition Eq.(\ref{BC})
by taking into account VL elasticity (collective pinning).  In the
AC experiment the electromagnetic fields produce additional
quasistatic uniform displacements ${\mathbf u}$ superimposed   on
the static random displacements induced by pinning. Because of
surface disorder the uniform displacement produces a random force
on vortices, which can be obtained from expansion of the random
pinning force $ {\mathbf f}({\mathbf r}_i)=-\varepsilon\varphi_0
{\mathbf\nabla }\zeta({\mathbf r}_i+{\mathbf u})$ with respect to
${\mathbf u}$: $\delta {\mathbf f}_m({\mathbf
r}_i)=-\varepsilon\varphi_0
 {\mathbf u}_n{\partial^2 \zeta({\mathbf r}_i)/{\partial  x_m
\partial  x_n}}$. However, the uniform displacement does not produce an
average force: $\langle\delta {\mathbf f}({\mathbf
r}_i)\rangle=0$. The fluctuating force produces fluctuating
displacements $\delta {\mathbf u}({\mathbf r}_i)$, which can be
found in the Fourier presentation where $\delta {\mathbf
u}({\mathbf k})=\delta {\mathbf f}({\mathbf k})/C({\mathbf k})$.
In contrast to the uniform displacement, the fluctuating
displacements $\delta {\mathbf u}({\mathbf r}_i)$ do produce an
average pinning force which  should be balanced by the uniform
 line-tension force:
\begin{equation}
{\partial {\mathbf u}_{\mathrm m}  \over \partial z} +
\left\langle \frac{\partial^2\zeta ({\mathbf{r}}_i)}{\partial
x_{\mathrm m}\partial x_{\mathrm n}}\;\delta{\mathbf u}_{\mathrm
n}({\mathbf r}_i)\right \rangle= 0\qquad.
       \label{rose}\end{equation}
Since $\delta{\mathbf u}$ is proportional to ${\mathbf u}$, we
arrive at the boundary condition Eq. (\ref{BC}) imposed on the
averaged, i.e., uniform displacement with slippage length given by
\begin{equation}
\frac{1}{l_S}\simeq  {\varepsilon\varphi_0 \over  4\pi^2 }
\sum_{{\mathbf Q}}\int {\mathbf dk} \,|{\mathbf k}+ {\mathbf Q}|^2
\left[ Q^2 - { ({\mathbf k\cdot Q})^2\over k^2}\right]  \times
{S_\zeta({\mathbf k}+ {\mathbf Q})\over C( k)}\quad.
\label{slippage_length}\end{equation}
 The same approximations as
in calculating $\left\langle{ ({\mathbf \nabla
u})^2}\right\rangle$ yield
\begin{equation}
{1\over l_S} \simeq{\varepsilon\varphi_0 a_0^2\over 8\pi } \int
_{0}^{2/a_0} {kdk \over C(k)}\int_0^\infty S_\zeta( Q) Q^5\,dQ
\approx\sqrt{\varepsilon B\over  C_{66}}{ 5! a_0\over 2 l_{0}^2}.
\label{lS}  \end{equation} Comparing with the expression $l_S=l_0
L_c/a_0$ we see that the size of the Larkin-Ovchinnikov domain is
$L_c \sim l_0a_0\sqrt{C_{66} /\varepsilon\varphi_0}$. In deriving
Eq. (\ref{lS}) we have used the perturbation theory, which is
valid until $L_c \gg a_0$, or $l_S \gg l_0$.

For low magnetic fields $B\!\ll\!B_{c2}$ one has $\varepsilon \sim
(\varphi_0/\mu_0\lambda^2)\ln(B_{c2}/B)$, $C_{66} \sim \varphi_0
B/\mu_0\lambda^2$ and according to Eq. (\ref{lS}) $l_S \propto
\sqrt{B/\ln(B_{c2}/B)}$. Then the surface pinning length $L_s
\propto [B/\ln(B_{c2}/B)]^ {3/2}$ grows with $B$ in qualitative
agreement with the experiment (Fig. \ref{pic_effect}). This is the
regime of collective pinning when $l_S > l_0$. At the same time
since $r_d \ll a_0$ the vortex lattice shear deformation remains
small according to Eq. (\ref{def}). For fields close to $B_{c2}$,
$\varepsilon \sim(B_{c2}-B)/\mu_0\kappa^2$, $C_{66} \sim
(B_{c2}-B)^2/\kappa^2$, and $r_d \sim a_0 \sim \xi$. Then Eqs.
(\ref{def}) and (\ref{lS}) yield $\left\langle{ ({\mathbf \nabla
u})^2}\right\rangle \approx (\xi^2/l_0^2)B_{c2}/(B_{c2}-B)$  and
 $l_S \approx (l_0^2/\xi) \sqrt{(B_{c2}-B)/B_{c2}}$. Thus $l_S$
decreases when $B$ approaches  $B_{c2}$ and for $B_{c2}-B <
B_{c2}\xi^2/l_0^2$ becomes smaller than $l_0$. This means that
surface pinning ceases to be collective and the crossover to
individual pinning occurs. At the same time, at $B_{c2}-B \sim
B_{c2}\xi^2/l_0^2$, the deformation $\left\langle{ ({\mathbf
\nabla u})^2}\right\rangle$ becomes of order unity, which means
that the crystalline order at the surface is destructed even at
short scales $\sim a_0$. We call this state {\em surface glass}.
Thus the crossover from collective to individual pinning is
accompanied by the crossover from a weakly disordered crystal to a
glass state  at the surface.

Still, this crossover cannot explain a fully developed PE. Despite
$l_S \propto \sqrt{B_{c2}-B}$ decreases at $B$ approaching to
$B_{c2}$, according to Eq. (\ref{two_mode}), $1/L_S$ continues to
decrease proportionally to $\sqrt{B_{c2}-B}$, whereas in the
experiment (Fig. \ref{pic_effect}) $1/L_S(B)$ increases on the
left of the peak. Nevertheless, the growth of the deformation
$\left\langle{ ({\mathbf \nabla u})^2}\right\rangle$, which
accompanies the decrease of $l_S$ eventually invalidates the
linear elasticity theory used above. Qualitatively this can be
corrected by introducing the renormalized deformation-dependent
shear modulus: $\tilde C_{66}=C_{66}(1  - \alpha \left\langle{
({\mathbf \nabla u})^2}\right\rangle) \approx C_{66}(1
-B/B_{pk})$. Here the field $B_{pk}$ corresponds to the
crystal-glass transition, where $\tilde C_{66}=0$, and $\alpha$ is
an unknown numerical factor, which could be close to 0.1 as in the
Lindemann criterion. Using renormalized modulus $\tilde C_{66}$ in
place of $C_{66}$ in Eq. (\ref{lS}) we obtain that $l_S$ [as well
as $L_S$, see Eq. (\ref{two_mode})] decreases proportionally to
$1/\sqrt{B_{pk}-B}$ in qualitative agreement with  experiment
(Fig. \ref{pic_effect}). On the right of the peak  pinning is
individual and $l_S \sim l_0$ does not depend on $B$, while $1/L_S
\propto (B_{c2}-B)$ decreases with  $B$.

The close relation between PE and vanishing of the shear modulus
of VL was suggested in the early studies of PE
\cite{Pippard69,Larkin}. The new feature of our scenario is that
at $B< B_{pk}$ the shear modulus  vanishes only at distances on
the order of the deformation penetration depth $1/p \propto
1/\sqrt{\tilde C_{66}}$ from the surface. Our scenario agrees with
 STM imaging of the vortex array by Troyanovski {\em et al.}
\cite{Troy}. They revealed that PE is accompanied by the disorder
onset on {\em  the surface} of a 2H-NbSe$_2$ sample, but they
related it with bulk pinning. In order to discriminate two
scenarios it would be useful to supplement the STM probing of the
vortex array at the surface by probing vortex arrangements in the
bulk.

In conclusion, we presented the experiment and the theory, which
support a new scenario for the peak effect based on competition
between vortex-lattice shear rigidity and weak surface disorder.
The peak is accompanied by a crossover from collective to
individual vortex pinning and from a weakly disordered crystal to
a glass state at the sample surface.  Beside its experimental
relevance, this mechanism offers an interesting paradigm for
elastic systems at the upper critical dimension for disorder.

 We thank F.R. Ladan and E. Lacaze for the
ion-beam etching and AFM measurements.  We acknowledge discussions
with B. Horowitz, T. Natterman and T. Giamarchi. This work was
funded by the French-Israel program Keshet and by the Israel
Academy of Sciences and Humanities. The Laboratoire Pierre Aigrain
is "unit\'{e} mixte de recherche" (UMR8551) of the Ecole Normale, the
CNRS, and the universities Paris 6 and Paris 7.

\pagebreak

\end{document}